# x

# Variations sur un thème : une approche reproductible

Nicolas Lambert[1], Timothée Giraud[1] et Ronan Ysebaert[1]
[1] *RIATE, CNRS, Paris, France*

## x.1. Introduction

Toute réalisation cartographique est affaire de choix. L'objectif de ce chapitre consistera à démontrer par l'exploration et la pratique que chaque carte est la résultante de décisions prises par le cartographe : sélection de l'information, choix graphiques, niveau de complexité méthodologique ; leur prise en considération permet d'affirmer une stratégie de communication pour une transmission efficace du message pour un public donné.

Dans ce chapitre, nous expliquons, exemples à l'appui, que les cartes sont autant des objets scientifiques par les techniques et le socle théorique sur lequel repose leur construction que des armes redoutables de communication. Pour appuyer cette démonstration, nous montrons à partir d'une même donnée géographique très conventionnelle, une grille de population à l'échelle mondiale, que la communication cartographique peut s'exprimer de multiples façons et renvoyer diverses formes de représentations. Nous souhaitons ainsi apporter la preuve qu'il n'existe pas de relation mécanique entre la donnée et son expression graphique et que toute carte résulte d'une intention. Car en réalité, plus que les aspects techniques et méthodologiques, c'est bien cette intention cartographique qui régit l'ensemble des décisions lors de la conception de la carte pour aboutir *in fine* à un document intelligible, organisé et communicable. ,

Pour cela, nous avons pris le parti d'ancrer notre démonstration dans une démarche traçable, transparente et reproductible. Chaque carte est ainsi associée aux codes sources R avec lequel elles ont été générées.

Une première partie traitera des enjeux généraux de la cartographie de communication. Après une courte introduction, nous aborderons à la fois la dimension scientifique et politique des cartes. Un point d'attention particulier sera porté aux aspects rhétoriques mobilisés dans les cartes thématiques, qui rendent d'autant plus nécessaire leur traçabilité. Puis, dans une seconde partie, nous proposerons la mise en pratique de ces concepts. À partir d'un même jeu de données portant sur la population mondiale, une série de 13 cartes sera présentée et discutée, chacune étant réalisée selon une intention et une méthodologie spécifique et véhiculant son propre message. Enfin, nous réunirons dans une dernière partie un certain nombre de procédés graphiques faisant écho à la thématique traitée, dans le but de décupler la puissance rhétorique des cartes.

## x.2. Chercher et communiquer avec les cartes

La carte est un puissant vecteur de communication pour transmettre une information sélectionnée, organisée et hiérarchisée sur une partie plus ou moins complexe du réel. Elle permet de visualiser d'un seul coup d'œil de vastes espaces non directement perceptibles. Aucun astronaute par exemple, aussi haut placé soit-il dans l'espace, ne peut embrasser du regard à un instant *t*, l'intégralité des pays du Monde. Et aucun plongeur ne peut observer l'étendue de la topographie de l'ensemble des fonds marins. La carte est donc un dispositif technique qui permet de montrer ce que nul œil ne peut voir (Jacob 1992) et ce qu'aucune photographie ne peut montrer. Aussi, pour faire rentrer l'entièreté du monde sur une feuille de papier ou un écran, celui-ci est mesuré, triangulé, numérisé, stocké, simplifié, et déformé par des opérations mathématiques. Puis, pour rendre compte d'un phénomène géographique particulier, des données thématiques peuvent être collectées, agrégées, transformées, associées et enfin mises en image sous la forme de signes graphiques.

### x.2.1. *La carte comme méthode scientifique*

La cartographie est un champ scientifique qui repose sur un socle méthodologique défini par des théories et des pratiques, et qui continue d'élaborer un « mode d'emploi » (Brunet 1987) du signifié, entre technique et médiation visuelle (Ferland 2000). Elle permet de fixer l'invisible et invite à réfléchir à ce qui nous entoure en nous extrayant des schémas de perception individuels. En ce sens, elle rend concret l'espace. Beaucoup plus qu'un simple assemblage de chiffres statistiques dans un tableau, la carte permet de faire émerger et de rendre visible des structures, des logiques et des dynamiques spatiales, qui sont autant de clefs d'explications et de compréhension du monde. On pense ici aux cartes de Cooper et Snow qui, en 1858, ont permis d'établir le lien entre l'épidémie de choléra qui sévissait à Londres et la contamination par l'eau (Cameron et Jones 1983 ; Koch et Denike 2009).

À travers cet exemple comme de bien d'autres, on voit que la carte (et par extension les systèmes d'information géographique et la cartographie en ligne) constitue une méthode rationnelle qui facilite l'accès à la connaissance en proposant une lecture spatiale. Pour le chercheur en sciences sociales, avec sa capacité de mise en forme et de modélisation, la carte est un outil largement utilisé pour explorer des données, tester des hypothèses et confronter des approches avec des pairs. L'utilisation de la cartographie participe en ce sens à la construction scientifique. Cet outil est aussi particulièrement adapté pour synthétiser et communiquer des connaissances et des résultats vers des publics non experts, mais en capacité de lire et comprendre ces documents graphiques.

### *x.2.2. Un outil éminemment politique*

Cette formalisation de l'espace qu'est la carte est rarement gratuite ou désintéressée (Lacoste 1976). Comme le souligne Bahoken (2020), le motif cartographié n'est pas muet, il est doté d'une signification qui traduit l'intention du concepteur (ou de l'auteur) d'une carte. La représentation des migrations est un bon exemple pour illustrer cette idée (figure 1). Entre les mains de l'agence Frontex[1], la carte sert à déterminer où les frontières extérieures de l'Union européenne sont les plus exposées à l'afflux de migrants, et en conséquence où elles sont à « protéger » par un déploiement de troupes. Cette vision est d'ailleurs souvent affirmée par des choix cartographiques très forts : les couleurs, la flèche, les mots (Houtum et Bueno Lacy 2020). Pour les ONG, les cartes servent au contraire à dénoncer l'absurdité et l'inefficacité des politiques migratoires, ainsi que des pratiques contraires aux droits humains les plus élémentaires. Entre les mains de dirigeants politiques nationalistes, elles peuvent être utilisées pour planifier et suivre l'évolution de la construction des murs à leurs frontières. Pour les activistes, les cartes revêtent une dimension tactique pour creuser des tunnels ou déterminer des passages discrets permettant d'échapper à la police lors du franchissement des frontières (Mogel et Bhagat 2007). Tantôt outil au service d'un pouvoir dominant, tantôt support pour le dénoncer ou le contourner, la carte peut se révéler être un objet pleinement politique (Stienne 2019).

---

[1] Frontex est une agence européenne dédiée à la gestion de la coopération opérationnelle aux frontières extérieures des États membres de l'Union européenne.

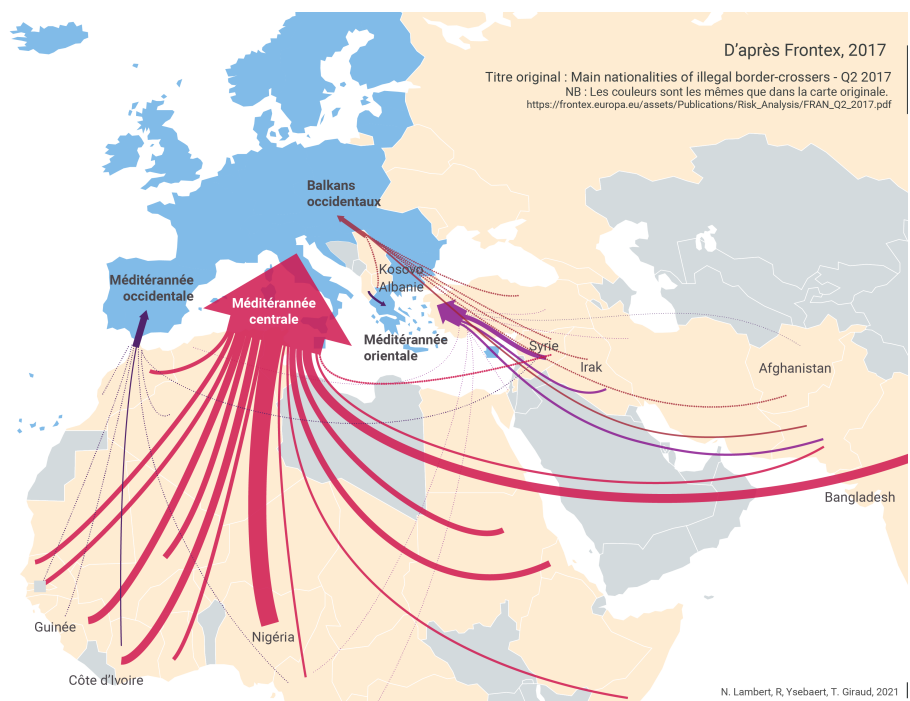

**Figure x.1.** *La rhétorique de l'invasion selon Frontex*

### x.2.3. *Cartographier c'est choisir*

La cartographie est donc une pratique profondément dialectique dans la mesure où elle permet de démontrer, de réfuter, d'argumenter et de raisonner. Elle est également à la fois intrinsèquement scientifique et objective, et tout aussi subjective et partiale. La subjectivité des cartes, que les géographes ont mis du temps à assumer en tant que telle, s'explique de différentes manières. Tout d'abord, aucune carte ne peut prétendre à elle seule représenter toute la complexité du monde réel. Selon Goodman (1972) « il n'existe aucune carte complètement adéquate car l'inadéquation est intrinsèque à la cartographie ». C'est un fait, la clarté exige sélection, simplification et généralisation. On ne peut pas tout mettre sur une carte. Une bonne carte sera donc celle qui simplifie, résume, hiérarchise, omet, ordonne, bref qui réalise une schématisation du réel indispensable à l'émergence des clefs de compréhension dans le cadre d'un sujet, d'un objectif particulier. La carte sert à « comprendre et chercher ce qui se passe derrière les apparences » (Pelletier 2013).

En raisonnant par l'absurde, de nombreux auteurs, de Carroll (1893) à Eco (1992) en passant par Borges (1946), ont digressé sur ce que serait une carte non simplifiée (Palsky 1999), à savoir une carte taille réelle, qui représenterait tout du monde et qui viendrait se superposer en tout point à lui. On le comprend, au-delà de l'impossibilité d'une telle production, il s'agirait là d'une entreprise bien inutile (voir à ce sujet le conte *la carte de l'Empire,* raconté par Chamussy en 1982). La simplification fait partie intégrante de la démarche cartographique. Certains auteurs iront même jusqu'à dire que les cartes mentent, qu'elles déforment la vérité pour aider l'utilisateur à voir ce que l'auteur veut qu'il voie (Monmonier 1991). En définitive, toute carte résulte d'une intention, une vision intellectuelle mais surtout de choix que le cartographe doit accepter, endosser, discuter et expliquer.

### x.2.4. *Le langage cartographique et sa rhétorique*

Afin de raconter et proposer des représentations du monde, les cartographes disposent d'un langage graphique varié qu'il est possible de mettre en œuvre avec une palette d'outils diversifiés, chacun ayant ses avantages et ses inconvénients. Toute carte porte en elle un propos qui s'exprime par des couleurs, des tracés, des mots, des exagérations et des omissions (Lambert et Zanin 2020). L'éventail de possibilités offert aux cartographes est large : sélection des données, choix du maillage et de l'espace géographique, transformations statistiques, transformations géométriques, choix graphiques, couleurs, habillage, mise en page, projection, généralisation, emprise, etc. Notons à ce stade, que s'il existe un lien fort entre la nature de la donnée utilisée et la façon de la représenter, il n'existe cependant pas de relation mécanique entre les deux. En d'autres termes, une même information peut donner lieu à une multiplicité de représentations (Zanin et Lambert 2012). Le langage cartographique rejoint donc ici l'art de la narration, de l'éloquence. Il y a les faits et la façon dont on les raconte. En cartographie, il y a les données et la façon dont on les met en carte.

Les concepts fondamentaux de la rhétorique, qui ont été définis par les philosophes grecs il y a plus de vingt siècles et notamment par Aristote, s'appliquent très bien à la cartographie contemporaine. Ils reposent sur trois piliers : *Logos*, *Pathos*, *Ethos* (figure 2). Le *Logos* fait appel à la raison, à l'intellect, au raisonnement logique. C'est le message en tant que tel, le contenu. Le *Pathos* fait référence quant à lui, aux affects, aux émotions, à l'imaginaire. En cartographie, cela consistera par exemple à choisir des tracés ronds et harmonieux plutôt que des angles aigus et abrupts, ou arbitrer entre le fait de privilégier des signes de petite taille pour minimiser l'importance d'un phénomène ou au contraire les grossir fortement, jusqu'à saturer graphiquement l'image, pour lui donner du poids. Enfin, *l'Ethos*, renvoie aux valeurs et à la crédibilité. Il détermine la confiance accordée au producteur de la carte, la renommée d'un chercheur, la réputation d'une institution, d'un journal ou la crédibilité des données représentées. Plus une carte a de *l'Ethos*, plus les arguments qui y sont déployés ont de l'autorité.

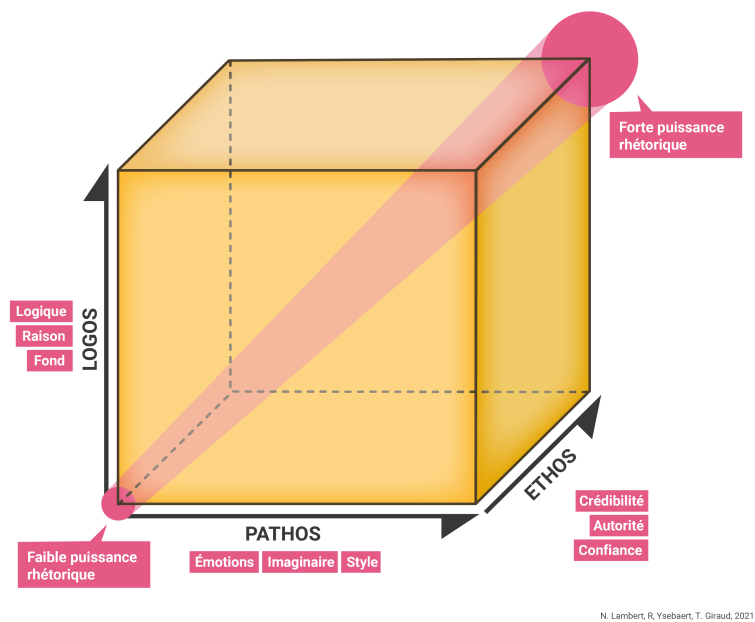

*Figure x.2. Le cube de la rhétorique cartographique*

Pour bien comprendre le rôle de la rhétorique en cartographie, les travaux de Bunge et notamment sa carte des nourrissons mordus par des rats dans les quartiers pauvres de Détroit en 1975 constituent un bon exemple (Popelard et Vannier 2009). En élaborant une telle carte (figure 3), Bunge ne se limite pas à représenter spatialement une simple donnée statistique, il en propose une vision. À travers un détail saillant et brutal, les morsures de rats sur des nourrissons, c'est en fait toute une réalité sociale qu'il révèle : insalubrité, précarité, pauvreté, inégalités, insécurité, etc. Or il s'agit là précisément d'une technique rhétorique bien connue : utiliser dans un récit ou un discours des détails si vifs qu'ils obligent ceux qui l'écoutent à reconstruire intérieurement l'intégralité de la scène. On est bien là dans l'art du récit. Bunge faisait d'ailleurs la distinction entre ce qu'il appelait *skeletal maps* qui représentent des informations statistiques et les *life maps* qui racontent la vie des gens, deux approches complémentaires selon lui (Bunge *et al.*1971).

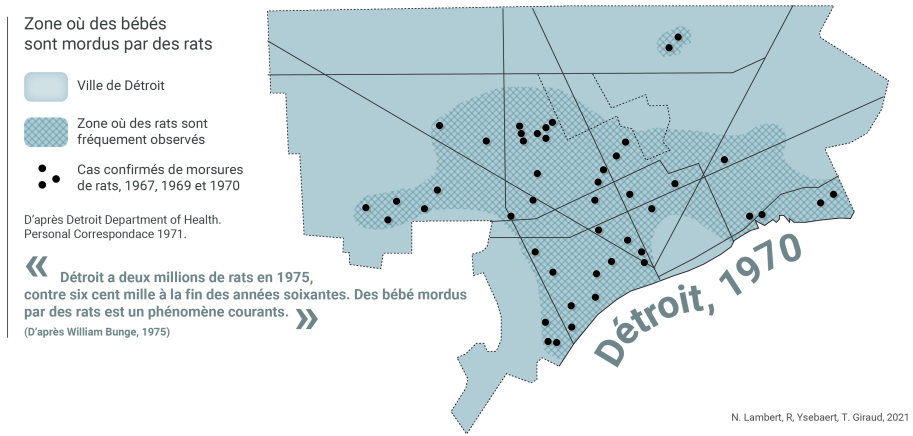

*Figure x.3. La géographie radicale de William Bunge*

En résumé, une carte sera d'autant plus efficace qu'elle sera éloquente et capable de captiver un public donné (la construction de la carte elle-même étant pensée en fonction du public à laquelle elle est destinée) pour lui transmettre un maximum d'information en un minimum de temps. Pour cela, l'art de la cartographie consistera à construire une image crédible, cohérente, parlante, où tous les éléments de la carte seront reliés entre eux selon une logique discursive qui parlera à la fois à la raison et aux affects (Migreurop et Clochard 2017).

### x.2.5. *Des transformations traçables et reproductibles*

Dans la mesure où les cartes « mentent » (le premier « mensonge », consistant à passer de la sphère au plan) et qu'elles relèvent de processus intentionnels de communication, qu'en est-il de leur scientificité ? Il sera toujours difficile, voire impossible, de démontrer le caractère objectif ou neutre d'une carte, raison pour laquelle il faut tout au moins être capable d'en garantir *a minima* l'honnêteté. Cela passe bien sûr par une volonté d'informer plutôt que de manipuler. Mais cela passe aussi par le fait de considérer la création cartographique comme la finalité d'un processus de transformation traçable (Tobler 1979), documenté. Rendre accessible les données, communiquer sur les traitements, partager les codes et expliquer les processus de mise en carte participent à la fois à rendre plus transparentes les pratiques cartographiques et à garantir une certaine honnêteté des cartographes comme de leurs productions.

Mais qu'en est-il dans nos pratiques effectives de cartographes ? Comment mettre en œuvre ces éléments théoriques de cadrage face à la réalité d'un jeu de données ? C'est l'objet de la prochaine section, où, avec pour support un jeu de données unique, une grille de population à l'échelle du monde, nous proposons une série de 13 cartes mobilisant sous des formes diversifiées la boîte à outils du cartographe (filtres, agrégations, discrétisations, transformations spatiales…) pour présenter (et communiquer) différemment sur la répartition de la population mondiale.

## x.3. Multireprésentation cartographique

Après une présentation des données utilisées et quelques éléments méthodologiques de cadrage, nous explorons et discutons ici les aspects discursifs liés aux modes de représentation cartographiques.

### x.3.1. *Principes de la démarche*

Nous proposons ici de démontrer, exemples à l'appui, en quoi et comment une carte est capable de dire beaucoup plus que la simple donnée brute qui a servi à sa construction. Pour cela, nous reprenons à notre compte l'exercice de style de Queneau (1947), déjà transposé à la cartographie sur le PIB des pays du monde (Lambert et Zanin 2016) et sur les personnes ayant péri en migration en Méditerranée (Giraud et Lambert 2019). L'objectif d'une telle démarche est de proposer une multiplicité de modes de représentation cartographiques construites sur des données identiques. Car comme un auteur de littérature, le cartographe doit faire des choix. Il doit à la fois sélectionner l'information, la trier, la hiérarchiser, et la sélectionner, dans le but de la résumer d'une façon intelligible et qui fait sens, et dans le même temps la représenter de manière claire et efficiente.

### x.3.2. *Mise en œuvre technique de la démarche*

Dans le cadre de cette démarche, nous avons choisi d'utiliser le langage de programmation et le logiciel libre R (R Core Team 2020) pour produire l'ensemble des cartes présentées dans ce chapitre. Ce choix technique se justifie par la qualité de l'écosystème spatial du langage (Bivand 2020), notamment en termes d'évolutivité comme de fonctionnalités. Un grand nombre d'extensions (les packages) permettent à la fois de mettre en œuvre un large ensemble d'analyses spatiales et de produire une grande variété de représentations cartographiques. De plus, la possibilité d'exécuter la quasi-totalité des chaînes de traitement (acquisition, mise en forme, traitement et représentation des données) au sein d'un même environnement permet d'éviter les ruptures logicielles et facilite *in fine* la reproductibilité de la démarche et des résultats (Giraud et Lambert 2017).

> IMPORTANT. Toutes les cartes présentées dans ce chapitre sont disponibles en ligne en haute définition accompagnée du code source qui a permis de les générer : https://rcarto.gitpages.huma-num.fr/multirepresentation. La mise à disposition des chaînes de traitement permet, dans une logique de transparence, d'accroître la confiance dans nos productions en permettant d'ouvrir la « boîte noire » de la fabrique cartographique (Nüst et Pebesma 2020).

La figure 4 décrit où se positionne notre démarche sur un spectre de la reproductibilité cartographique (Peng 2011). Dans notre démarche, la mise en forme des données, leur traitement et leur représentation cartographique ont été effectués grâce à des programmes en langage R. Ces programmes utilisent les librairies R de référence pour mettre en forme et manipuler des données spatiales (*raster*, *sf*), puis les cartographier (*cartography*). Certaines librairies plus spécifiques sont également mobilisées pour des traitements plus avancés (calcul de potentiel de Stewart avec *SpatialPosition*) ou des modélisations spécifiques (*tanaka*, *linemap*). Cette partie du travail est donc parfaitement traçable et reproductible, comme en atteste le site Web adossé à ce chapitre, qui restitue l'intégralité du code utilisé pour générer les représentations cartographiques présentées ici.

*A contrario*, il est souvent nécessaire de recourir à des logiciels de création graphique vectorielle pour parachever leur mise en forme : la présentation finale de ces cartes (hiérarchisation des textes, positionnement des étiquettes, conception des légendes, etc.), qui relève davantage de la conception graphique a été effectuée avec un logiciel de DAO[2], où les techniques peuvent certes être documentées mais ne sont pas reproductibles.

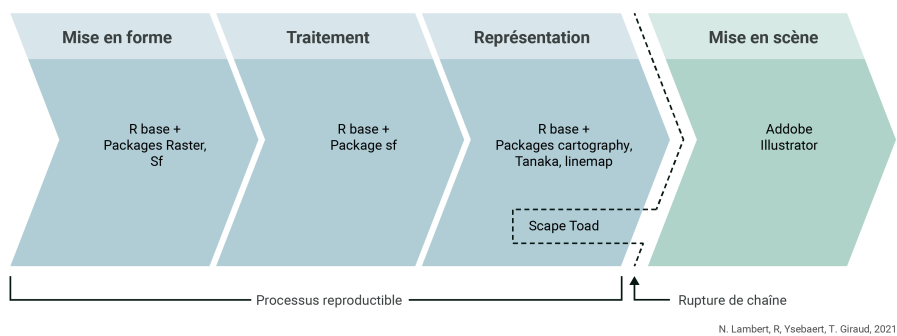

***Figure x.4.*** *Spectre de la reproductibilité en cartographie*

---

[2] Dessin Assisté par Ordinateur

**x.3.3. *Les données de la démarche***

Toutes les représentations cartographiques qui vont suivre ont pour origine un seul et même jeu de données, *the Gridded Population of the World* (GPW)[3]. Parmi la dizaine d'indicateurs mis à disposition par le CIESIN, nous retiendrons le plus simple et le plus actuel : la population totale en 2020.

Les indicateurs démographiques proposés par le CIESIN sont issus d'une très importante collecte de données réalisée au niveau administratif le plus fin possible pour tous les pays du monde à partir des recensements nationaux entre 2005 et 2014. Ces données statistiques sont ensuite appariées aux géométries de référence de chacun des pays, et estimées pour l'année 2020 en suivant les tendances observées sur les deux derniers recensements et les estimations des Nations Unies, pour finalement être ventilées dans une modélisation raster (GeoTIFF) à différents niveau de granularité. La résolution la plus fine disponible correspond à 30 secondes (environ 1 km à l'équateur) et la plus large à 60 minutes (environ 110 km à l'équateur). Pour cet exercice, c'est la résolution de 30 minutes qui a été retenue[4]. À ce niveau de résolution spatiale, on dénombre 70 123 carreaux recouvrant des terres émergées dans le Monde (figure 5).

| Population 2020 des carreaux | Occurrences | Fréquence (%) |
|---|---|---|
| Moins d'1 habitant | 11 075 | 15,8 |
| 1 à 10 habitants | 2 184 | 3,1 |
| 10 à 100 | 7 247 | 10,4 |
| 100 à 1 000 | 8 973 | 12,8 |
| 1 000 à 10 000 | 13 550 | 19,3 |
| 10 000 à 100 000 | 15 361 | 21,9 |
| 100 000 à 1 000 000 | 10 028 | 14,3 |
| 1 000 000 à 23 065 668 | 1 705 | 2,4 |
| **Somme** | **70 123** | **100** |

*Figure x.5. Dénombrement par intervalle de population (échelle logarithmique) des cellules de la grille du CIESIN (résolution = 30 minutes)*

---

[3] Le GPW est un quadrillage (raster) produit par le CIESIN (*Center for International Earth Science Information Network* de l'université de Colombia) qui modélise la distribution de la population humaine (dénombrements et densités).
[4] Du fait des distorsions introduites par les projections, la résolution du maillage équivaut à des carreaux de 55 km de côté au niveau de l'équateur, 35 km aux latitudes intermédiaires (Paris), et à moins de 10 km au nord du Groenland.

Cette grande disparité de distribution des populations au sein des carreaux s'observe également en termes de fréquence cumulée (figure 6). On constate ainsi que 50 % de la population mondiale (plus de 4 milliards d'individus) se concentrent dans seulement 2,4 % des 70 123 cellules du jeu de données. À l'inverse, 50 % des cellules du jeu de données cumulent moins de 14 millions d'habitants (0,2 % de la population mondiale).

La distribution de cette variable statistique, très dissymétrique à gauche (la médiane équivaut à un carreau peuplé de 2 708 habitants, contre 110 600 pour la moyenne) révèle à ce stade des opportunités intéressantes pour le cartographe qui souhaite transcrire graphiquement les spécificités de ce jeu de données. Comment restituer les concentrations démographiques maximales contenues dans certains carreaux ou certaines portions de l'espace mondial ? Comment évoquer les « vides », qui occupent la majeure partie de la planète ?

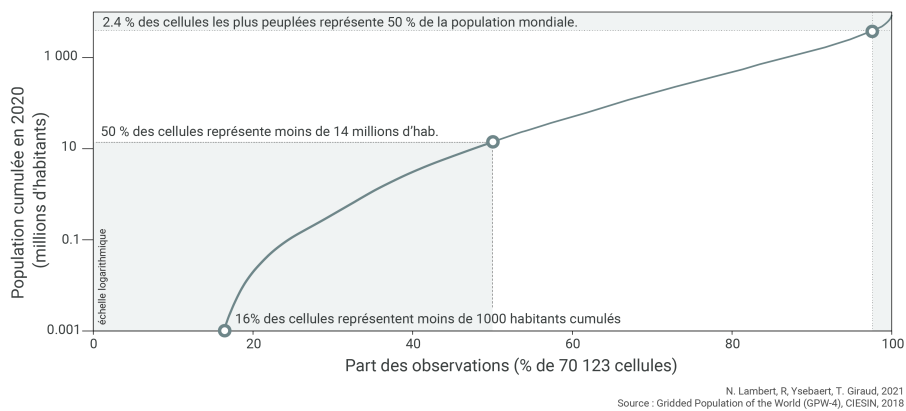

**Figure x.6.** *Fréquence cumulée*

Cette courte description du jeu de données utilisé rappelle aussi quelques éléments importants à avoir en tête avant de se lancer dans toute production cartographique. Tout comme les cartes, les données sont des construits sociotechniques (Chrisman, 2005). Leur collecte, leur traitement, leur mise à jour et leur distribution ont un coût, des objectifs politiques, sociaux ou économiques et il existe toujours un biais sur le fait qu'elles traduisent une réalité objective. En ce sens, l'un des cas d'étude le plus saisissant concernant des données de population est apporté par l'exemple du recensement nigérian où après plusieurs recensements effectués dans des conditions chaotiques, ce pays, le plus peuplé d'Afrique, a du jour au lendemain perdu 30 millions d'habitants en 1991 (Lévy et Omoluabi 1992).

Le cartographe, grand utilisateur et « manipulateur » de données, doit donc idéalement s'en servir en connaissance de cause sur l'autorité, l'administration ou l'organisation qui les ont produites, et mesurer les conséquences qu'elles peuvent induire et impliquer dans la production cartographique. Car les données mobilisées, avec des sources mal maîtrisées ou renseignées peuvent indubitablement introduire le mensonge cartographique et desservir la crédibilité du message.

### x.3.4. *Constante N° 1 : choix d'une projection cartographique*

Nous l'avons dit plus haut, toutes les cartes mentent, le premier mensonge étant de représenter un monde sphérique en trois dimensions sur une surface plane en deux dimensions. Cette critique n'est d'ailleurs pas nouvelle. Au XIX[e] siècle, Élisée Reclus considérait déjà qu'enseigner la géographie sur des images fausses était un non-sens total (Reclus 1903). Pourtant, du fait même de leur capacité à déformer, les projections cartographiques, dans leurs diversités, sont des armes précieuses entre les mains des cartographes car la façon de déformer le monde véhicule en soi une vision du monde. Au-delà des enjeux de précisions géodésiques, nombre de projections cartographiques sont conçues en pratique pour le message qu'elles véhiculent (figure 7). Le message de chaque carte, est toujours le fruit d'une composition complexe associant un contenu (la donnée statistique représentée) et un contenant, le « décor » sur lequel elle est ancrée (Rekacewicz 2014). Chaque carte relève ainsi d'une mise en scène qui n'est pas sans rappeler le monde du théâtre.

La projection de Gall-Peters par exemple, en montrant la superficie réelle des continents au détriment de leur forme, vise à réhabiliter les pays du sud écrasés sur les représentations cartographiques classiques par un Nord dominant. La projection polaire, utilisée pour le logo des Nations Unies, qui prend le parti de ne mettre aucun pays au centre, est née au lendemain de la Seconde Guerre mondiale pour instaurer une volonté de paix et d'égalité. Quant à la projection cartographique de l'océanographe Athelstan Spilhauss, très originale, elle ne vise rien d'autre que d'attirer l'attention sur l'importance des mers et des océans à la surface de la terre. Bien plus que de simples supports pour l'information géographique, les différentes projections cartographiques sont des vecteurs d'éléments discursifs, qui portent en eux une narration, et font partie intégrante du message cartographique.

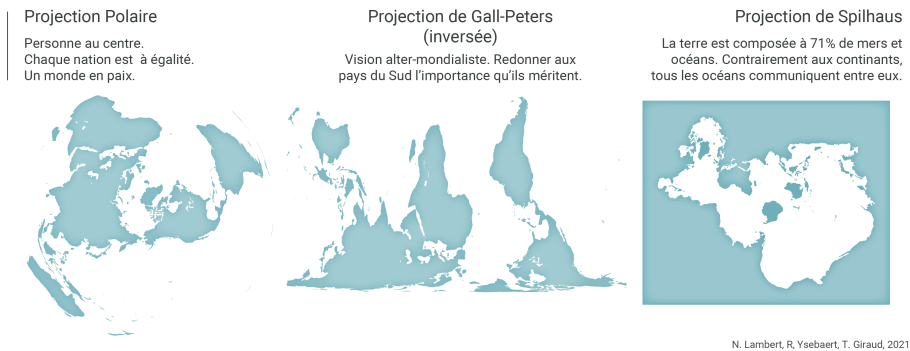

*Figure x.7. À chaque projection son message*

Bien que conscients de l'étendue des possibilités rhétoriques des projections cartographiques, nous avons situé celles-ci en dehors de notre spectre d'analyse. Dans la démonstration à suivre, nous raisonnerons à projection et donc « décor » constant, celui d'un monde déformé par une projection cylindrique équidistante centrée classiquement sur l'Europe. Ce choix d'une représentation du monde habituelle, la plus neutre possible a comme objectif de mettre l'emphase sur les représentations cartographiques des données statistiques.

### x.3.5. *Constante N° 2 : choix d'une échelle cartographique*

Pour les mêmes raisons, nous travaillerons à échelle constante, celle du monde. Le découpage géographique et l'échelle d'analyse exercent une influence importante sur les résultats modélisés. Chaque échelle d'analyse ou d'observation apporte ses clés de lecture (figure 8). « Tantôt, il faut regarder la terre au microscope et tantôt du haut d'un satellite […] La réalité apparaît différente selon l'échelle des cartes, selon les niveaux d'analyse » (Lacoste 1976). Dans l'exercice de style cartographique présenté ici, nous travaillerons donc à l'échelle mondiale uniquement, à projection cartographique et à niveau de granularité géographique constant, en nous appuyant sur les données mises à disposition par le CIESIN à une résolution de 30 minutes. Dans certains cas, ces données pourront être agrégées dans des carroyages de différentes tailles ou des maillages administratifs.

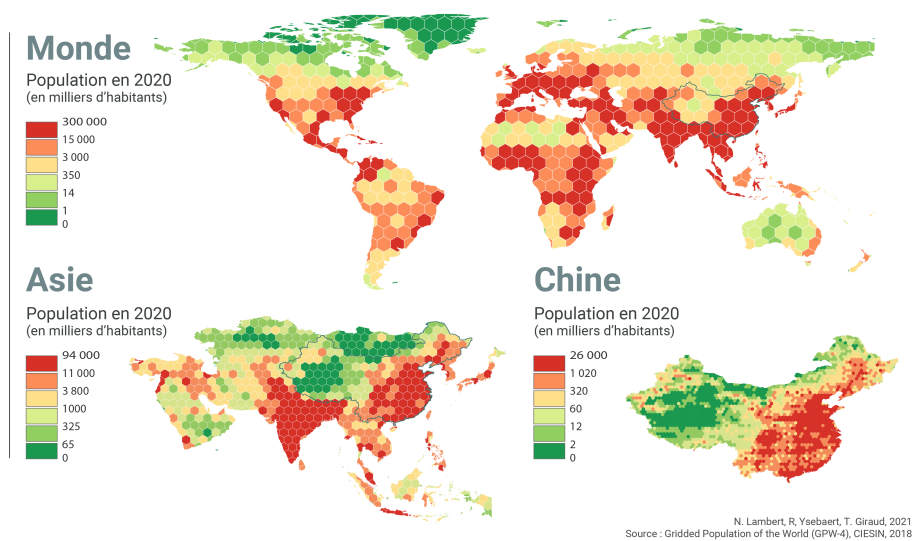

*Figure x.8.* La « réalité » apparaît différente selon l'échelle

### x.3.6. *Le discours des données*

Le choix de représenter des données relatives ou des données absolues est un choix lourd de sens qui s'évalue au regard de la thématique considérée. Comment estimer par exemple la richesse d'une nation ? En considérant le nombre de dollars en circulation dans un territoire donné ou bien en appréciant cette même richesse au regard de la population concernée ? Niveau de vie ou puissance économique, tout dépend de ce que l'on souhaite montrer. Cartographier des données absolues ou des données relatives, c'est renvoyer à des réalités différentes et surtout communiquer des messages distincts, divergents voire contradictoires. Passer de l'absolu au relatif peut en effet constituer un moyen de minimiser et relativiser un phénomène, ou au contraire lui donner une dimension plus sociale. Passer du relatif à l'absolu c'est rappeler que, dans les rapports de force, c'est toujours le nombre qui compte.

Appliquée à la population mondiale, la représentation cartographique des données absolues par des signes proportionnels (figure 9), renvoie au poids démographique des États tandis que la représentation des densités de population par aplats de couleurs (figure 10) focalise le message cartographique sur l'inégale répartition des humains sur Terre et mettra ainsi plutôt en exergue les petits pays densément peuplés (Monaco, Malte, Bahreïn).

Sur les cartes en proportion, la sémiologie graphique permet de se « libérer du cadre » imposé par le fond de carte pour donner toute leur place aux signes. En passant d'une implantation zonale (les pays) à une figuration ponctuelle (les signes), le cartographe « ruse » en imposant à la carte des figurés qui débordent plus ou moins largement du strict lieu qu'elles désignent ; ou au contraire, ne le couvrent que partiellement (Brunet 1987), permettant ainsi d'exprimer au mieux les proportions représentées. *A contrario*, les cartes choroplèthes, construites sur des données relatives discrétisées, ne permettent quant à elles que d'exprimer une relation hiérarchique entre les lieux, le choix des bornes de classes pouvant d'ailleurs fortement influencer le message. Si ces deux méthodes sont utiles et complémentaires pour analyser et représenter un espace géographique, elles délivrent, par construction, des messages de nature radicalement différente.

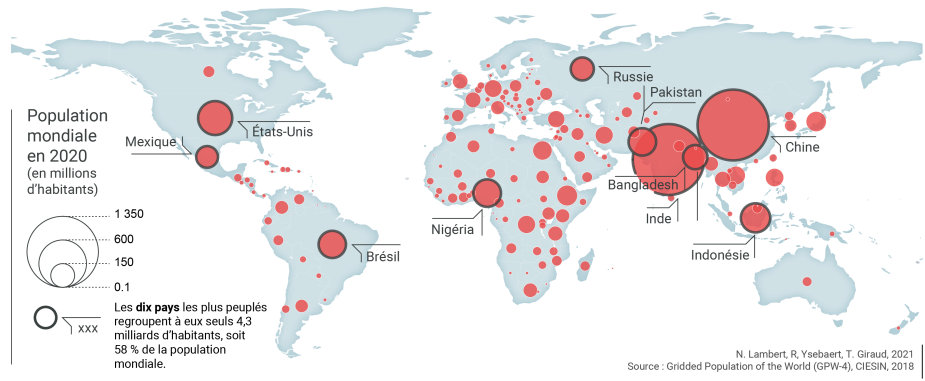

**Figure x.9.** *L'inégale puissance démographique des États-Nations*

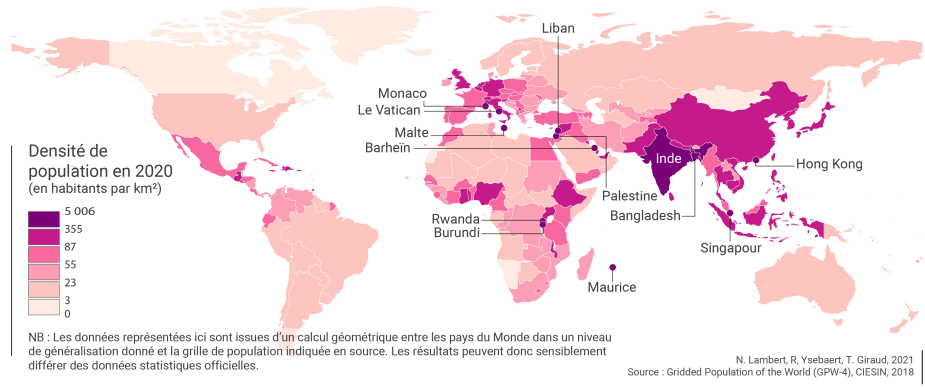

*Figure x.10.* Densités de population

Cette idée de puissance peut aussi être véhiculée dans un certain nombre de transformations cartographiques (Cauvin 1998), parmi lesquelles les transformations cartographiques de poids (une catégorie d'anamorphose) qui vont jusqu'à s'affranchir de la géographie physique et transforment les contours des objets géographiques qui servent de support à la représentation cartographique. Ici, les déformations cartographiques effectuées selon la méthode développée par Gastner et Newman en 2004 (Gastner et Newman 2004) et implémentée dans le logiciel *ScapeToad* (Andrieu *et al.,* 2008), modifient la taille de chaque pixel en fonction de la valeur de population qui lui est associée, avant de les regrouper par pays, pour reformer un maillage territorial alternatif (figure 11).

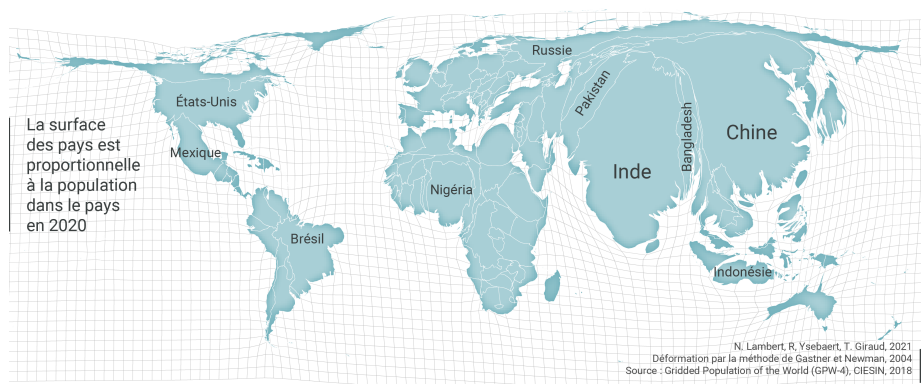

*Figure x.11.* Distorsion démographique

Sur ce type de cartes, les grandes puissances démographiques prennent place au-delà de leurs propres « frontières » tandis que les pays les moins peuplés se contractent jusqu'à parfois disparaître de la surface de la carte. L'utilisation de ce nouveau fond de carte débarrassé de sa « réalité » géographique produit ainsi une véritable rhétorique du rapport de force avec des gagnants et des perdants.

### x.3.7. *Sortir du carcan géométrique*

Loin de ces images qui exacerbent la grandeur et l'omniprésence des États-Nations, et alors même que les données statistiques sont souvent fournies par ces

États eux-mêmes et que l'un des corrélats historiques des États-Nations est la statistique (Grataloup 2009), d'autres représentations cartographiques visent plutôt à s'affranchir des découpages territoriaux. Certaines méthodes de transformation permettent par exemple de désagréger ou d'interpoler les données pour les représenter différemment des maillages de collecte et des découpages administratifs. Si cela a un intérêt méthodologique majeur pour surmonter les difficultés statistiques et cognitives liées à l'hétérogénéité des maillages (Openshow 1979 ; Grasland *et al.* 2006), cela permet aussi de simplifier les structures spatiales en montrant l'espace géographique selon des logiques « naturelles » ou continues (Commenges *et al.* 2015). En d'autres termes, cela permet de proposer des visions du Monde non plus internationales mais mondiales (Grataloup 2018 ; Capdepuy 2020).

La méthode de désagrégation des données la plus ancienne est probablement la carte par points, la première réalisation connue étant celle d'Armand Joseph Frère de Montizon en 1830 sur laquelle un point représentant 10 000 personnes (Palsky 1984). Ce type de cartes très pédagogiques est très présent dans les manuels scolaires pour décrire les foyers de peuplement. Les cartes par points donnent l'image d'une représentation au plus proche du terrain et traduisent la répartition des êtres humains à l'échelle de la planète d'une manière aussi objective que possible (figure 12). Cette impression est d'autant plus forte que la localisation des points est précise à l'intérieur des États (près des côtes, des grandes villes, des axes de communications, etc.). Mais n'oublions pas qu'il s'agit d'une image construite de toute pièce et que les procédés techniques sous-jacents sont complexes. Ici par exemple, les points sont répartis à l'intérieur d'un carroyage selon une méthode aléatoire. Plus la maille de calcul est fine, plus le rendu visuel donne l'impression d'une image proche du réel.

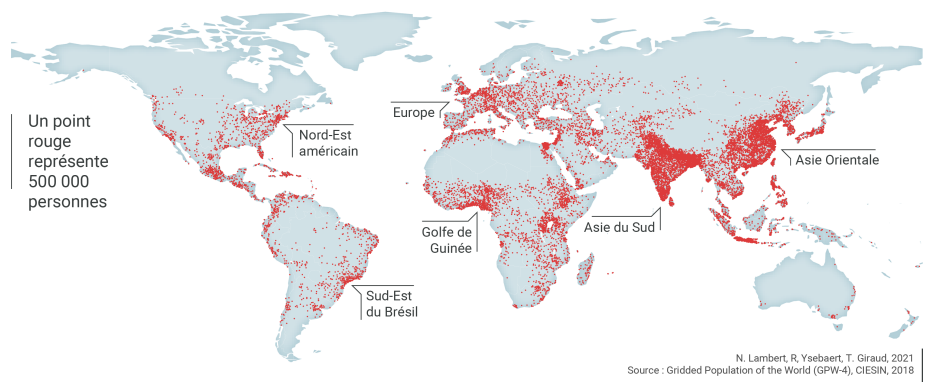

N. Lambert, R, Ysebaert, T. Giraud, 2021
Source : Gridded Population of the World (GPW-4), CIESIN, 2018

*Figure x.12. La population mondiale en 2020*

Une variante de cette représentation de la population mondiale par points consiste à répartir des signes proportionnels (souvent des cercles) régulièrement à la surface de la carte en d'en faire varier la surface en fonction des données sous-jacentes. Pour plus de lisibilité, ces signes peuvent être écartés les uns des autres (Dorling 1996) jusqu'à sortir des mailles initiales. Il s'agit d'une technique intéressante pour éviter la superposition des figurés cartographiques les uns par rapport aux autres (cartes en cercles proportionnels classiques) et montrer de façon synthétique l'ensemble des masses en présence sur un espace géographique donné. De façon symbolique, tout comme les anamorphoses, cette représentation peut évoquer l'idée du débordement (figure 13). Dans notre exemple, des puissances régionales (et non plus étatiques) émergent (l'Asie du Sud, l'Asie orientale, le Golfe de Guinée) tandis qu'apparaissent, en creux, des espaces vides.

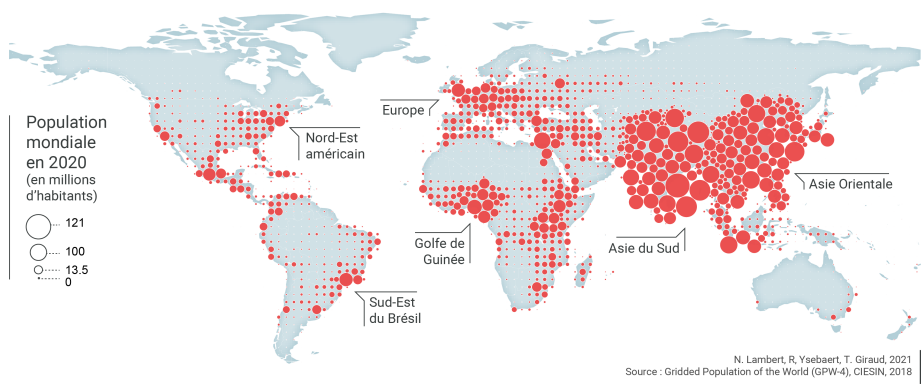

*Figure x.13. Débordement démographique*

Pour s'affranchir des frontières, les méthodes de lissage et d'interpolation spatiale apportent d'autres perspectives (Stewart 1942). En se basant sur le calcul des masses dans un voisinage géographique donné dont l'importance décroît avec la distance (calcul de potentiels[5]) puis en extrayant des isolignes et des isosurfaces de valeurs comparables, ces méthodes permettent de dessiner un monde de continuums indépendant des frontières, bien loin des logiques discontinues des cartes des États nation (figure 14). Produire une telle carte, c'est proposer un monde de gradients, sans ruptures, où les données s'étalent d'un bout à l'autre du globe terrestre. Mais d'autres peuvent y voir aussi (ESPON 2012) une manière indirecte de discuter de « reliefs démographiques », où la chaîne démographique du Penjab côtoie les pics côtiers de la mer de Chine, où le front de cuesta maghrébin plonge dans la vaste dépression saharienne ; ou encore l'immense atoll latino-américain laissant entrevoir quelques résurgences démographiques au centre de la forêt amazonienne.

---

[5] Ce calcul repose sur une fonction mathématique décroissante avec la distance utilisée pour évaluer la masse de population dans un voisinage géographique (200 kilomètres ici). Les paramètres fondamentaux utilisés dans ce calcul (fonction, impédance de la fonction, et distance) sont légendés dans la carte ci-dessous. Leur bon usage, fondé sur des hypothèses d'interaction entre des lieux, est essentiel pour une réalisation maitrisée de ce type de représentation cartographique.

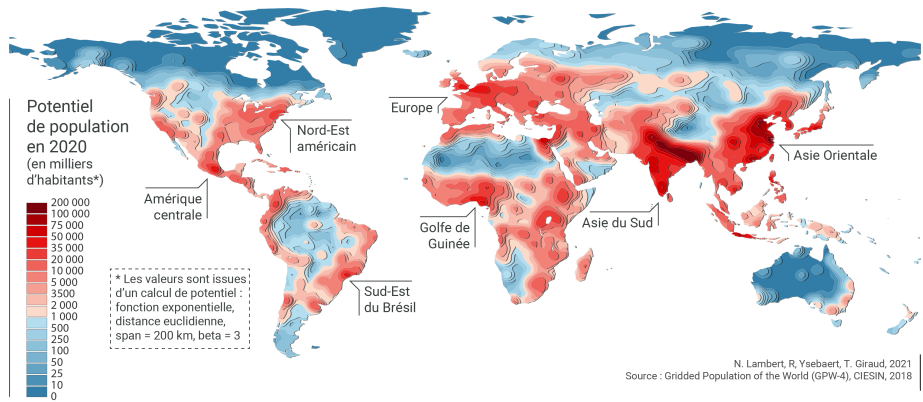

*Figure x.14. Un Monde sans frontières*

### x.3.8. *La question du public*

Quelle que soit la méthode retenue, aucune carte ne peut être construite sans prendre en compte le public auquel celle-ci est destinée. Enfants, adultes, spécialistes, décideurs, scientifiques, nous ne sommes pas habitués aux mêmes images cartographiques ni à la même complexité des discours sous-jacents.

Ainsi, dans une volonté de sensibiliser voire de convaincre le plus de monde, on voit apparaître sur Internet et les réseaux sociaux un certain nombre de cartes simplifiées à l'extrême qui ont pour but d'être comprises en un temps extrêmement court (au détriment de la complexité réelle des phénomènes représentés). Sur ce nouveau genre cartographique (Schweikart et Domnick, 2013), on est plus proche de l'infographie, avec des chiffres clés mis en valeur que des constructions classiques de cartographie thématique où priment les structures spatiales. Sur ces représentations, le message doit compris en un coup d'œil. Pas de temps de la réflexion, ce qui compte c'est le partage en masse, l'audience, les réactions, et surtout la viralité de la carte sur la toile (Robinson 2018).

Résumer la complexité d'un phénomène par un chiffre ou une image percutante n'est pas une mince affaire, mais s'avère particulièrement efficace quand la démarche est bien menée (l'avancement de l'urbanisation en terrains de football, l'empreinte écologique exprimée en nombre de planètes…). Synthétiser un phénomène inextricable et complexe en une « punchline » permet de toucher le plus grand nombre. Cette logique vaut aussi en cartographie, nombre de cartes circulent et sont largement relayées et commentées sur le web et les réseaux sociaux car leur message est insolite et/ou compris en un laps de temps extrêmement bref. Les deux cartes ci-dessous (figures 15 et 16) en sont des exemples. Elles n'ont pas d'intérêt géographique majeur, mais ce type de représentation répond à une attente toujours forte du grand public qui désire disposer d'éléments factuels pour comprendre le monde, représentés, cependant, de manière simple et efficace. Encore faut-il, qu'en simplifiant à l'extrême on n'introduise pas une information erronée.

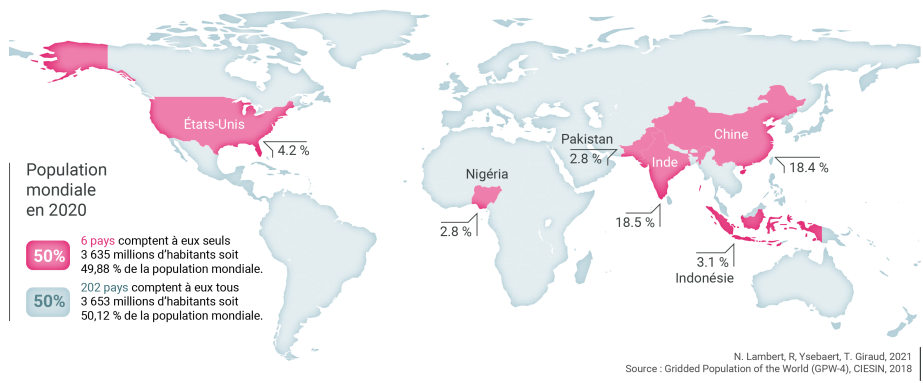

*Figure x.15. La moitié de la population vit dans seulement 6 pays*

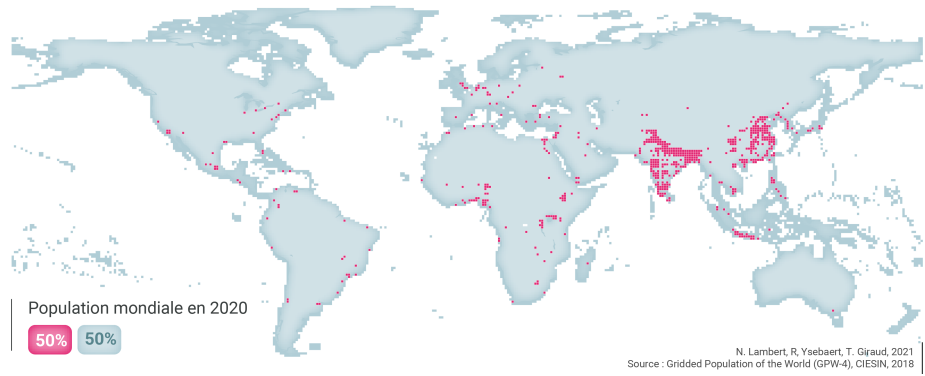

**Figure x.16.** *La moitié de la population mondiale vit sur 3 % de la surface terrestre*

### x.3.9. *Quand la sémiologie évoque une thématique*

Au fur et à mesure qu'on explore la diversité des cartes réalisées autour d'un même jeu de données, il apparaît que les possibilités cartographiques pour exprimer et communiquer des idées voire des intentions sont presque infinies. Parfois, afin de rendre cohérente la narration cartographique, on cherchera même à utiliser un procédé technique qui fait écho à la thématique traitée. La carte ci-dessous (figure 17) reprend à son compte le principe de la première loi de la géographie, avancé par Tobler (1970), pour qui « Tout interagit avec tout, mais deux objets proches ont plus de chances de le faire que deux objets éloignés ».

C'est cette idée qui a été reprise ici par un système de lignes qui relie les lieux peuplés de plus de 3 millions d'habitants situés à moins de 500 km les uns des autres (selon une distance euclidienne). Ce seuil, bien qu'arbitraire, permet de rendre compte d'un monde interconnecté où deux lieux proches sont susceptibles d'interagir davantage que deux lieux éloignés. Par ailleurs, les lignes sont un moyen habile de souligner et rendre visibles les régions plutôt denses et proches les unes des autres. Notons que cette carte aurait pu être produite avec un seuil de distances différentes et une autre agrégation spatiale plus ou moins fine.

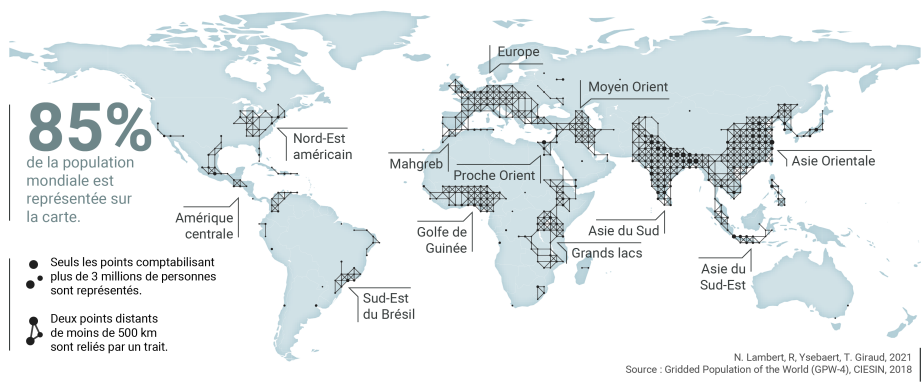

*Figure x.17. Tout interagit avec tout...*

On le voit, lier la méthode de représentation et le message peut s'avérer très efficace. Alors comment exprimer le fait que plus de la moitié de la population mondiale vit aujourd'hui en ville ? La solution proposée ici a été de symboliser les données en trois dimensions et de travailler sur la hauteur (Franklin et Lewis 1978). Les foyers de populations urbaines sont ainsi représentés par des mailles hexagonales extrudées symbolisant des immeubles dont la hauteur varie en fonction du nombre d'habitants (figure 18). La forte urbanisation de certaines zones de monde se matérialise ainsi sur la carte par une série de colonnes urbaines (*skyline*) où se mêlent à la fois hauteur et densité des immeubles.

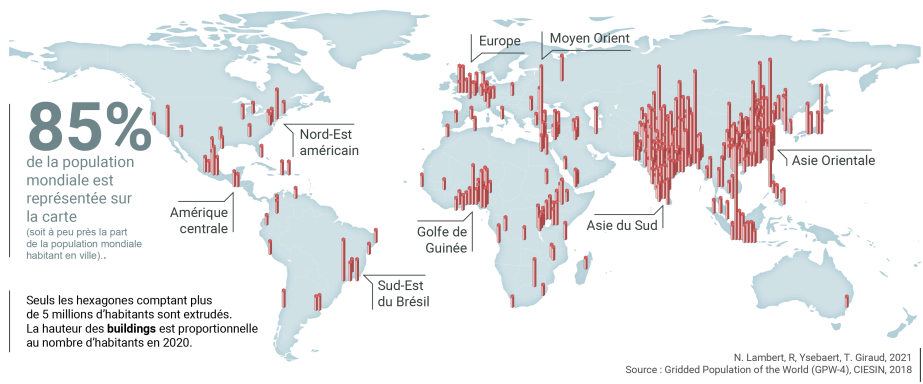

*Figure x.18. Homo Urbanus*

Par le jeu de la sémiologie graphique, de la statistique, des géotraitements… il y a mille et une façons de faire parler une donnée géographique et la transformer en discours graphique. Au delà des aspects strictement liés à la sémiologique graphique en tant que telle, ne sélectionner que les données à proximité des mers et océans permet par exemple de révéler que près de la moitié de la population mondiale vit près d'une côte (figure 19). En focalisant sur cet aspect, la carte prend alors une dimension politique, puisque selon le GIEC, la montée des eaux pourrait atteindre plus d'un mètre d'ici la fin du siècle.

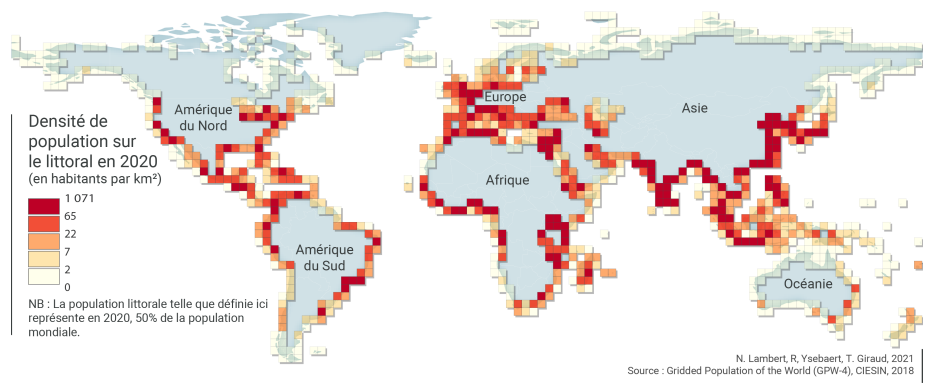

*Figure x.19 L'humanité en proie à la montée des océans*

De même, focaliser non plus sur les lieux peuplés en valeur absolue mais sur les lieux relativement peuplés au regard de leur voisinage (la résolution et la définition du voisinage étant bien sûr paramétrables), permet de montrer une carte où les zones mises en évidence ne sont pas forcément celles où les habitants sont les plus nombreux ni même où la densité de population est la plus forte (Grasland 1999 ; de Ruffray *et al.* 2011 ; Dubois *et al.* 2007). Une telle carte (figure 20), qu'on pourrait caractériser de « localiste », donne en effet une tout autre image du Monde, une image « glocale »[6] qui met en avant les potentialités locales.

Cette représentation propose une autre vision où les pôles isolés émergent et côtoient les grandes dorsales métropolitaines mondialisées. Elle permet aussi de mettre en lumière un réseau d'agglomérations inhabituel, lié spécifiquement aux fronts pionniers et routes historiques, que ce soit en Russie (le long du transsibérien), en Amazonie (le long de l'Amazone en passant par Manaus) ou à l'ouest de la Chine (route de la Soie).

---

[6] Le terme glocal est un mot valise qui signifie à la fois local et global. Il a été popularisé dans le monde anglophone dans les années 1990 par le sociologue Roland Robertson.

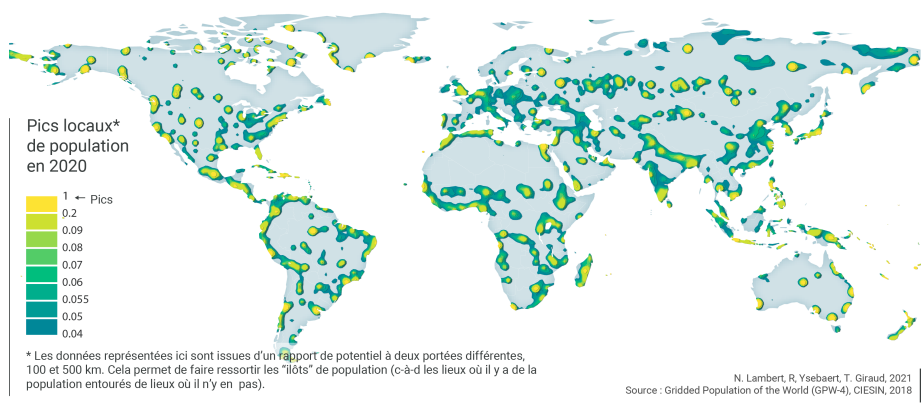

***Figure x.20.** Glocalisation*

Enfin, les cartes peuvent aussi être des invitations au rêve et à l'aventure. Les blancs sur les cartes ont d'ailleurs toujours fait rêver. Les terres inconnues ouvrent des perspectives d'aventure dont sont friands nombre de géographes. Aussi, cette carte des vides démographiques (figure 21) renvoie à cette idée. En axant la représentation sur des terres lointaines et inconnues (Tiberghien, 2007), elle invite à l'aventure. Évidemment, d'autres représentations auraient été possibles pour visualiser cette thématique à l'image des anticartogrammes qui consistent à représenter de façon inversée les pleins et les vides (Poncet 2011).

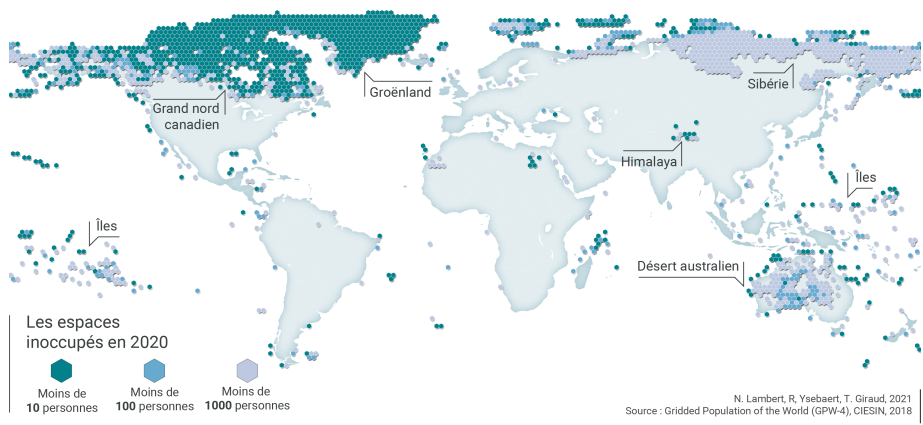

*Figure x.21.* Terrae Incognitae

## x.4. Conclusion

Cet exercice de multireprésentations cartographiques matérialisé par 13 cartes diversifiées illustre bien qu'à partir d'une même information, à projection, échelle et emprise constante, il est possible de construire un vaste éventail de propositions cartographiques, chacune d'elles étant bâtie selon sa propre méthode et véhiculant son propre message. À partir d'un jeu de données des populations mondiales, nous avons ainsi tantôt mis en lumière des masses et des hiérarchies, tantôt des concentrations spatiales (pics, valeurs exceptionnelles) et finalement des continuums et ruptures dans l'espace.

Une possible synthèse de cette palette de représentations spatiales consiste à croiser les modes de construction et les objectifs de communication pour révéler les intentions cartographiques (figure 22). Nous concevons ici « l'intention du cartographe » comme les choix méthodologiques et graphiques inhérents à la construction cartographique dans le but de communiquer un message vers un public donné. En fonction des modes de représentation, le lecteur n'acquerra pas les mêmes connaissances sur la répartition de la population mondiale.

Ces cartes peuvent ainsi être pédagogiques et accessibles : l'information géographique a été filtrée, épurée et contextualisée pour focaliser sur certains aspects du jeu de données. Elles sont à même également d'être « conventionnelles » et de mobiliser des variables visuelles caractéristiques de la sémiologie bertinienne. Mais elles sont susceptibles aussi de reposer sur des distorsions géographiques qui permettent d'appuyer graphiquement la démarche proposée par le cartographe. Elles ont enfin la possibilité de reposer sur des méthodes issues de l'analyse spatiale ou de la statistique, qui permettent de transformer la donnée originale faisant ainsi émerger des structures et des organisations spatiales spécifiques. Ce type de représentation cartographique sera nécessairement plus complexe à appréhender pour un public peu enclin à comprendre le socle méthodologique sur lequel leur construction repose, d'où l'intérêt de documenter et de partager les chaînes de traitements. Quoi qu'il en soit, l'éventail des possibilités graphiques émises par un cartographe pour communiquer à un public donné ce qui lui paraît important sera *a priori* d'autant plus large que sa maîtrise et son recul sur les méthodes (quel procédé graphique ou statistique pour faire quoi ?) et outils mobilisés (comment réaliser la carte ?) sera important.

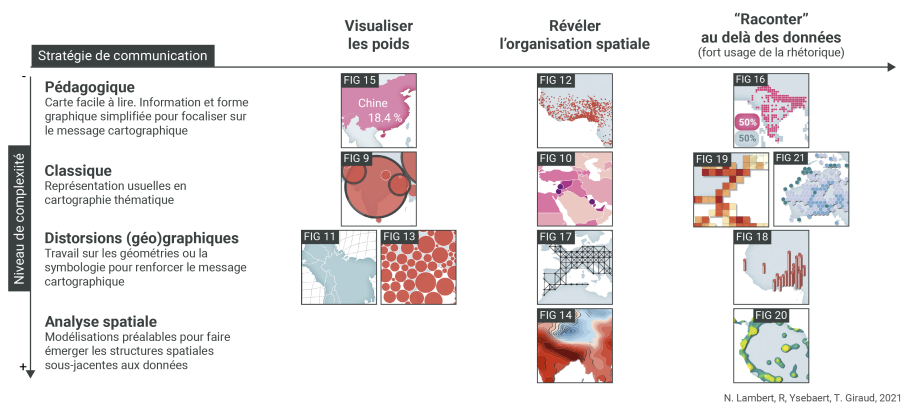

**Figure x.22.** *Synthèse de l'exercice de multireprésentation*

Au-delà de cette proposition de synthèse forcément perfectible, nous avons avant tout voulu faire la démonstration qu'il n'y a pas de relation mécanique et automatique entre les données géographiques et leur traduction cartographique, et qu'au delà les choix techniques, toute carte résulte avant tout d'une intention ; d'une stratégie de communication. Nous ne prétendons donc pas, à l'issue de cet exercice de style, avoir épuisé le sujet, au contraire. C'est bien en réalité tout un éventail de représentations que permettent la statistique et l'analyse spatiale. En transformant les données par la prise en compte de la distance, des surfaces, de la topologie, des maillages sous-jacents, la carte peut prendre des formes extrêmement diversifiées (continuums, ruptures, représentations multiscalaires, etc.) pour véhiculer des messages variés. Une des clefs de l'originalité et de l'efficacité des cartes ne se trouve donc pas uniquement dans la sémiotique des données géographiques mais aussi au niveau de la préparation et de l'analyse de ces données, en amont, qui sous-entend l'intention cartographique. Faire une carte de communication ne se limite pas uniquement à la « simple » mise en forme de données spatiales mais renvoie aujourd'hui à des compétences larges en termes de traitement des données. Cela souligne la forte dimension méthodologique intimement associée à la cartographie thématique. Cela est d'autant plus vrai au regard de l'instrumentation contemporaine, à l'image du langage R et de sa multitude de packages qui offrent de nombreuses possibilités en perpétuelle évolution tant pour la représentation cartographique que pour l'analyse spatiale.

D'autres facettes cartographiques auraient en ce sens pu être explorées et construites. On pense par exemple à l'exploitation de la troisième dimension (Grasland et Madelin 2001) ou aux possibilités d'interactivité et d'animation rendues possibles aujourd'hui par la cartographie en ligne. Nous soutenons ainsi l'idée que les choix de représentation vont bien au-delà de ce qui a été montré ici. Aucune carte ne résumera et n'épuisera à elle seule une thématique, ou même une donnée. C'est au contraire à travers une multiplicité de représentations spatiales à confronter, à discuter, à critiquer qu'il est possible de saisir la complexité d'un espace géographique, ainsi que ses contradictions.

Pour terminer, il ne faut pas perdre de vue que la capacité de séduction des cartes est grande et que leur esthétique frise parfois avec le monde de l'art, comme le rappelle cette dernière carte (figure 23) très largement inspirée de celle imaginée par Tobler en 1970. La séduction des cartes, une véritable arme de guerre en termes de communication, voire de manipulation, implique pour les cartographes de toujours réfléchir, expérimenter, repenser et se questionner sur leurs productions.

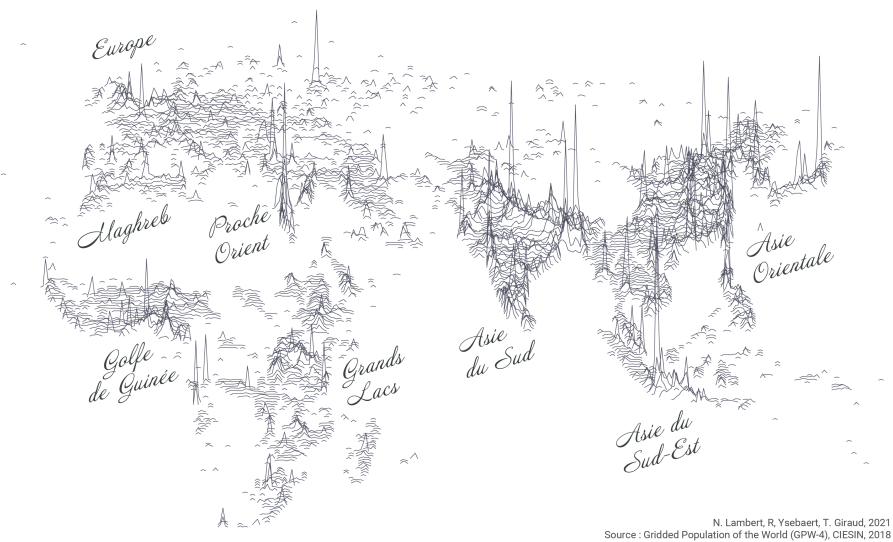

*Figure x.23.* L'art est dans la cARTe

**x.5. Bibliographie**